\documentclass{article}

\usepackage[preprint]{neurips_2026}

\usepackage[utf8]{inputenc} % allow utf-8 input
\usepackage[T1]{fontenc}    % use 8-bit T1 fonts
\usepackage{hyperref}       % hyperlinks
\usepackage{url}            % simple URL typesetting
\usepackage{booktabs}       % professional-quality tables
\usepackage{amsfonts}       % blackboard math symbols
\usepackage{nicefrac}       % compact symbols for 1/2, etc.
\usepackage{microtype}      % microtypography
\usepackage{xcolor}         % colors
\usepackage{graphicx}
\usepackage{amsmath, amssymb, amsthm}
\usepackage{makecell}
\usepackage{multirow}
\usepackage{changes}
\usepackage{algorithm}
\usepackage[noEnd]{algpseudocodex}
\usepackage{bm}
\usepackage{bbm}

\newcommand{\ket}[1]{|#1 \rangle}

\newcommand{\Mc}[1]{\mathcal{#1}}
\newcommand{\mb}[1]{\mathbb{#1}}
\newcommand{\bt}{\bm{\theta}}

\newcommand{\bh}{\bm{h}}
\newcommand{\bk}{\bm{k}}

\newcommand{\bQ}{\bm{Q}}
\newcommand{\bW}{\bm{W}}

\newcommand{\bracks}[1]{\{#1\}} 

\definechangesauthor[name={Ankit}, color=orange]{ak}

\theoremstyle{definition}
\newtheorem{definition}{Definition}[section]

\usepackage{changes}
\definechangesauthor[name={joey}, color=red]{jl}

\title{CO-MAP: A Reinforcement Learning Approach to the Qubit Allocation Problem}

\author{%
  Ankit Kulshrestha \\
  Fujitsu Research of America\\
  Santa Clara, CA - 95054 \\
  \texttt{akulshrestha@fujitsu.com}\\
  \And 
  \And
  Xiaoyuan Liu\\
  Fujitsu Research of America\\
  Santa Clara, CA - 95054 \\
  \texttt{xliu@fujitsu.com}\\
}

\begin{document}

\maketitle

\begin{abstract}
A quantum compiler is a critical piece in the quantum computing pipeline since it allows an abstract quantum circuit to be run on a physical quantum computer. One extremely important subproblem in quantum compilation is the generation of a logical to physical qubit mapping. Typically in quantum compilers this step is either implemented as a random or a heuristic based assignment that aims to minimize additional (SWAP) gate overhead in the quantum circuit. 

In this paper, we present an alternative approach to solving the qubit mapping problem. Specifically, we formulate the qubit mapping problem with a combinatorial optimization (CO) objective. We then present a method to find a solution to the CO problem by training a reinforcement learning (RL) policy. We also propose a local search based post-processing algorithm to further reduce the overhead. Our results show a dramatic improvement over conventional techniques in reducing the number of SWAPs. On different real world datasets like MQTBench and Queko circuits, our trained policy achieves a \textbf{65-85\%} reduction in SWAP overhead when compared to existing quantum compilers. 

% In this paper, we take a different view of the qubit mapping problem and define it in terms of a Combinatorial Optimization (CO) problem. We then present a reinforcement learning (RL) method that learns a policy that ``solves'' the CO problem. Our results show a dramatic \textbf{65-85\%} improvement in the solution quality over known commercial quantum compiler pipelines. Our method does not depend on a particular topology or advance knowledge of quantum circuits and requires a single pass over the given quantum circuit. Our problem formulation and results show that qubit mapping is an understudied problem in literature and AI techniques have a vast potential in solving the problem without manually defined heuristics.

\end{abstract}

\section{Introduction}

Quantum computing is touted to the be the next frontier in the field of computer science. The promise of so-called ``quantum-advantage" has already led to the formulation of several forward looking policies including discussions around post-quantum cryptography. As quantum computing devices continue to grow in terms of physical qubits and logical error suppression, the need for a quantum compiler to grow in lock-step with these advances is not only desirable - it is essential to  the success of quantum computing. 

% \ak{state this first sentence more cleanly}
Quantum compilers~\cite{booth2012quantum, chong2017programming} perform a central role in quantum computing since they enable an abstract quantum program to be translated into instructions that can be run on actual quantum hardware. There is a great variance in quantum hardware technology (e.g. IBM, Fujitsu have superconducting qubit architectures~\cite{kim2023evidence}, while IonQ has ion trap~\cite{monroe2021ionq} architecture) and connectivity. There are several sub stages in quantum compilation but in this work we focus on one of the earliest stages in compilation - layout generation and optimization. In quantum computing jargon, this problem is also known as qubit mapping or allocation problem~\cite{siraichi2018qubit}. Informally, this stage of the quantum compilation pipeline is responsible for coming up with a mapping from program qubits (i.e. the qubits that the quantum programmer thinks should interact) to physical qubits while respecting the connectivity constraints on the quantum device. 

The general attitude towards qubit mapping is that it is a necessary but unimportant first step towards the eventual stage of \emph{qubit routing}. Thus, most innovations in literature arise from finding algorithms to efficiently route qubits instead of finding good layouts for the given quantum program. However, we argue that finding a good mapping can significantly reduce the time spent in qubit routing and enhance the performance of the compiler directly. The oversight resulting from the established attitude towards this problem motivated us to study it in more detail. 

Our study revealed that most qubit mapping algorithms are heuristical in nature. The heuristics themselves are human designed. This very fact renders their generalization to be in question. Thus, in this work we formulate qubit mapping differently than in literature and propose a reinforcement learning (RL) algorithm that is able to \emph{learn} general heuristics for qubit mapping and outperforms manually tuned algorithms. In brief, we make the following contributions:
\begin{itemize}
    \item We formulate the qubit mapping problem in terms of a CO problem and propose a RL method that learns to solve the problem.
    \item We further demonstrate that our policy network learns \emph{data-driven} heuristics that generalize well beyond the training set and perform exceptionally well on real-world quantum circuits. 
    \item We introduce a lightweight post-processing algorithm that uses the solution produced by the trained policy to further improve the solution. The post-processing algorithm adds minimal processing time while delivering \textbf{51-88\%} improvement over the RL solution across different datasets. 
\end{itemize}

\begin{figure}
    \centering
    \includegraphics[width=\linewidth]{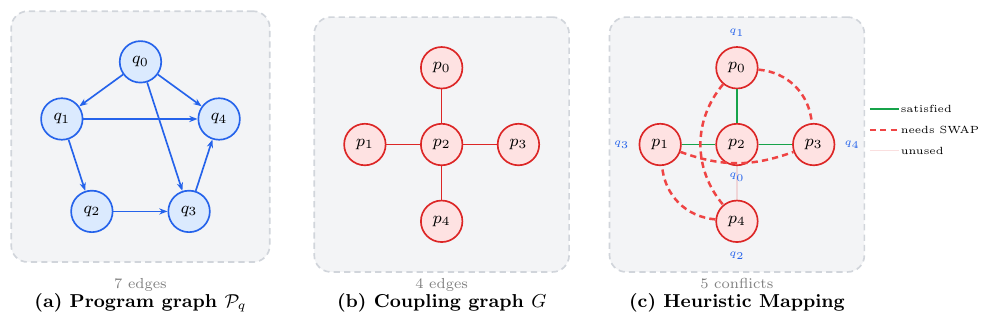}
    \caption{Qubit allocation with a given program and device graph. A heuristic mapping places the highest degree logical node on the ``hub'' of the device graph and allocates the neighbors on the surrounding qubits. Despite the heuristics, the circuit requires insertion of four SWAP gates to manage allocation. This illustrates that no algorithm cannot find perfect mapping from arbitrary program graphs to a given topology.}
    \label{fig:qm_example}
\end{figure}

\section{Related Work}

There is a rich body of work in quantum compilation that deals with the topic of qubit mapping and routing. Siraichi~\emph{et al}~\cite{siraichi2018qubit} define the problem of qubit allocation in terms of boolean satisfiability and provide a heuristic algorithm for mapping.  IBM's SABRE~\cite{yolcu2019learning, javadi2024quantum} proposes a search algorithm that iteratively finds mapping that reduces SWAPs. The Quilc~\cite{smith2020open} compiler proposes to solve qubit allocation stage (called ``addressing'' in the paper) by doing a topological sort on the input quantum circuit qubits and iteratively assigning them to the physical device. Liu~\emph{et al.}~\cite{liu2022leveraging} propose a heuristic-driven quadratic unconstrained binary optimization(QUBO) formulation for solving the qubit mapping. Recently, Molavi~\emph{et al}~\cite{molavi2026generating} proposed a different approach in which they \emph{generate} a quantum compiler given a quantum program and hardware by filling an abstract representation with the correct parameters. 

There has been a growing interest in applying RL techniques to the qubit mapping problem. For instance, Kremer~\emph{et al}~\cite{kremer2024practical} propose to model the qubit routing problem as a sequential decision problem and solve it using off-the-shelf policy gradient algorithms. Huang~\emph{et al}~\cite{huang2022reinforcement} propose to model qubit mapping  as an instance of sequence-to-sequence problem which they solve by using Transformer~\cite{vaswani2017attention} architecture. A related approach that is similar to ours in spirit is proposed by~\cite{russo2024attention} where they propose to use CO formulation for a multi-core quantum device architecture and train a policy that learns an assignment from a quantum circuit to a particular executable core. Each core is assumed to have all-to-all connectivity. In contrast, we focus on the much harder single core qubit mapping problem with the objective of reducing overall SWAPS for \emph{any} connectivity pattern in a single pass.

Our proposed approach is built over existing literature on using RL methods for CO problems~\cite{bengio2021machine}. Vinyals~\emph{et al}~\cite{vinyals2015pointer} proposed the original PointerNet architecture that was successively refined to solve various standard CO problems like TSP, CVRP etc.~\cite{bello2016neural}. Recently, Kool~\emph{et al}~\cite{kool2018attention} proposed an attention model formulation for these problems and it has inspired a fresh research into this area~\cite{yang2022graph,grinsztajn2023winner}. While this work is not an algorithm to solve CO problems in general, we hope to innovate quantum-aware RL methods for CO problems in a future work.

\section{Preliminaries}
\textbf{Quantum Circuits}: In this work, we define a program qubit as a state $\ket{\psi} = \alpha\ket{0} + \beta\ket{1}; \alpha,\beta \in \mb{C}$. This program qubit can only be assigned to one physical qubit. This is different from ``logical qubits'' which can occupy multiple physical qubits. A quantum circuit $\bm{U}$ acts on $\ket{\psi}$ to produce an output state $\ket{\phi}$. This output state is then \emph{measured} w.r.t a traceless Hermitian matrix $O$ (called an ``observable'') to produce information that can be processed by a classical computer. It is typical for quantum circuits to be represented as layers of same operations repeated upto a finite depth i.e. $\bm{U} = \prod_{\ell=1}^{L}U_{\ell}$. In this stage of compilation we assume that each layer consists of transpiled circuit such that $U_{\ell}^i = e^{-iW}V$ where $W$ is a non parameterized Hermitian matrix and $V$ represents the entanglement between various qubits in the layer. We refer an interested reader to excellent references on this subject~\cite{nielsen2010quantum, cerezo2021variational} for more information.

% Within a layer one may have gates that are parameterized i.e. $U_\ell^i = e^{-i\theta W}$ or $U_\ell^i = e^{-iW}$. Here $W$ is a non-parameterized Hermitian and $\theta$ is a rotation parameter
\textbf{Quantum Devices and Program Graphs}: Physical realization of qubits obey certain constraints owing to the topology of the quantum device. An abstract representation of this topology is called a \emph{\textbf{coupling graph}}. Formally, a coupling graph is a graph $G=(V_Q, \Mc{E}_Q)$ where $V_Q = \bracks{Q_1, Q_2, \dots Q_N}$ is the set of all physical qubits and $(Q_i, Q_j) \in \Mc{E}_Q$ is an edge indicating a physical connection between $i^{th}$ and $j^{th}$ qubit. Similarly, entanglement operations in an abstract quantum circuit can be represented as a directed \emph{\textbf{program graph}} $\Mc{P}_q = (V_q, \Mc{E}_q)$ where $V_q = \bracks{q_1, q_2, \dots q_n}$ is the set of all participating program qubits and $(q_i, q_j) \in \Mc{E}_q$ represents a two qubit operation between $i^{th}$ and $j^{th}$ program qubits. 

% For the purposes of layout mapping, we are not interested in single qubit rotation and transformations in a given quantum circuit. We are more interested in the entanglement operations since this defines the sequence and order of qubit interaction in a given quantum circuit. The abstract representation of this entanglement operation in a quantum circuit is called a \emph{\textbf{program graph}}. 

We define the qubit allocation problem as follows:
% Let $\Mc{Q} = \{Q_0, Q_1 \dots Q_P\}$ be the set of all physical qubits available on a quantum device and $G_Q$ be the coupling map that provides the interaction between those qubits. Additionally, let $\Mc{P}_q = \{q_0, q_1, \dots q_n \}$ denote a \emph{program graph} which describes the interaction between logical qubits in a quantum circuit.
\begin{definition}{\textbf{Qubit Allocation Problem}:}\label{def:qalloc}
Given a quantum circuit represented as a program graph $\Mc{P}_q$ and a quantum device represented by a coupling graph $G_Q$, the qubit allocation problem is to find an assignment $\Psi = \{q_i \mapsto Q_j\}_{i=1\dots n}^{j=1\dots N}$ such that all interacting pairs of program qubits $(q_i, q_j) \in \Mc{E}_q$ occupy adjacent positions on $Q_k, Q_l \in V_Q$. 
\end{definition}
Alternative versions of Definition~\ref{def:qalloc} define the problem in terms of boolean satisfiability~\cite{siraichi2018qubit}. Regardless of the definition, this problem is NP-Complete. In other words, given a program graph $\Mc{P}_g$ and a coupling graph $G_Q$, it is impossible to find a \emph{perfect} mapping $\Psi: V_q \mapsto V_Q$ from a set of program qubits $q$ to set of physical qubits $Q$ in polynomial time (Figure~\ref{fig:qm_example}). We \emph{can} however propose circuit transformations~\cite{siraichi2018qubit} to introduce additional gates that can allow program qubits to be moved according to the topology in $G_Q$. One such type of gate is the SWAP gate represents a permutation $\pi:\Mc{P}_q \mapsto \Mc{P}'_q$ such that the $q_i$ acts in the place of $q_j$ and vice versa. The qubit mapping problem can then be defined as a CO problem that minimizes the number of SWAPs inserted into the circuit:
\begin{align}\label{eq:co_qm}
    \min_{\bm{x}} \quad & \sum_{(i,j) \in \Mc{E}_q} \sum_{p=1}^{N} \sum_{k=1}^{N} 
        d(p,k)\, x_{i,p}\, x_{j,k} \\
    \text{s.t.} \quad 
        & \sum_{p=1}^{N} x_{i,p} = 1, 
            \quad \forall i \in \{1,\dots,n\} \label{eq:assign} \\
        & \sum_{i=1}^{n} x_{i,p} \leq 1, 
            \quad \forall p \in \{1,\dots,N\} \label{eq:capacity} \\
        & x_{i,p} \in \{0,1\}, 
            \quad \forall i \in \{1,\dots,n\},\; p \in \{1,\dots,N\} \notag
\end{align}

% \begin{align}\label{eq:co_qm}
%     \min \sum_{(i,j)\in \Mc{E}_q}\sum_{p=1}^N\sum_{k=1}^N d(p,k) x_{i, p} x_{j,k}\\\nonumber
%     s.t \sum_{p=1}^N x_{i,p} &= 1\ \forall i \in \{1, \dots n\}\\\nonumber 
%    \sum_{i=1}^n x_{i, p} \leq 1\  \forall p \in \{1, \dots N\}\\\nonumber
%     x_{i, p} \in \bracks{0, 1}
% \end{align}

In Equation~\ref{eq:co_qm}, we assume $N \geq n$, so that the quantum program can be executed on the underlying quantum device. The number of SWAPs to be inserted naively is considered to be $2*d(p,k)$ where $d(p,k)$ is the physical distance between $Q_p, Q_k \in V_Q$. The factor of $2$ arises due to the cost of applying an additional back SWAP so that the subsequent mapping is not affected.  The solution to the variable $x_{i,p} = 1$ represents that we place program qubit $i$ to physical qubit $p$, and 0 otherwise.

\section{Neural-CO for Qubit Mapping} 
% \ak{Rewrite the entire section more concisely and without too many subsections.}
% \ak{Additionally, details that are not strictly essential to the main paper can be referred to in the Appendix}

In this section we provide an overview of our method. We first begin by describing the observation and action space in our RL formulation. We then describe the architecture of our policy network followed by a light weight post-processing algorithm to further enhance the results.
 
\subsection{Reinforcement Learning Formulation}
% We wish to learn a policy $\pi_{\bt}(\Psi | \Mc{P}_q, G_Q)$ that represents the mapping mapping $\Psi: \bq \mapsto \bQ$. Following existing work in this area~\cite{} we define the problem as a sequential partially-observable MDP (POMDP) $P(\Mc{O}, \Mc{A}, \Mc{R}, \gamma)$ with the following variables.
The RL formulation for the CO objective is defined in terms of a model-free MDP (MDP) $P(\Mc{S}, \Mc{A}, \Mc{R}, \gamma)$ with the following definitions: 

\textbf{State Space}: The state space $s_t \in \Mc{S}$ consists of the following variables:
% consists of the input program graph $\Mc{P}_q$, the coupling graph $G_Q$. For a given $\Mc{P}_q$ we fix the order of logical qubits to be placed. We further augment the state space to include the following additional time dependent variables:
\begin{enumerate}
    \item $\Mc{P}_q$: The program graph describing the logical qubit entanglement pattern in the given quantum circuit.

    \item $G_Q$: The coupling graph describing the physical connectivity of the quantum device. 
    
    \item Allocation $\psi(t) \in \mb{R}^n$ tracks the partial mapping until timestep $t$.

    \item Distance matrix $\bm{D} \in \mb{R}^{N \times N}$ is the distance matrix representing distance between all physical qubits given $G_Q$.

    % \item Physical Qubits Used $\hat{Q}_t \in \mb{R}^n$ tracks the allocated physical qubits up until timestep $t$. 

    \item Current logical $c_t$ tracks the index of current logical qubit under consideration. The terminal state occurs when $c_T = n-1$. 
\end{enumerate}

\textbf{Action Space}: At a given decoding step we define the feasible action set $\Mc{A}(s_t) = \bQ \setminus \bracks{a_0, a_1, \dots a_{t' < t}}$. The action set is setup in this way because the constraints in Equation~\ref{eq:co_qm} prevent us from assigning more than one logical qubit to a single physical qubit. The decoder outputs logits over the entire $\bQ$ but we compute a masked policy:
\begin{equation}\label{eq:masked_qubit_policy}
    \pi_{\bt}(a | s_t) = \frac{\exp(f_{\bt}(s_t, a))\cdot\bm{1}[a\in \Mc{A}(s_t)]}{\sum_{a' \in \Mc{A}(s_t)} \exp(f_{\bt}(s_t, a'))}
\end{equation}

Where $f_{\bt}(s_t, a)$ is the logit output from the policy network for the given state $s_t$ and feasible action $a$.

\textbf{Reward Function}: In our formulation we do not provide immediate reward to the policy network after updating mapping from $\psi(t) \rightarrow \psi(t+1)$. Instead, once the terminal state for the given $\Mc{P}_q$ is reached, we compute the reward as:

\begin{equation}\label{eq:reward_func}
    R(\psi_T) = - \sum_{(i,j)\in \Mc{E}_q} \sum_{p=1}^N\sum_{k=1}^N r(p,k) x_{i, p} x_{j, k}
\end{equation}

Where $r(p,k) = 2d(p,k)$ is the SWAP count to excecute a two qubit operation between program qubit $q_i$ and $q_j$ after placing $q_i \mapsto Q_p$ and $q_j \mapsto Q_k$. The reward function computes the number of SWAPs produced under the current layout while assuming that there is no subsequent routing stage. The negation occurs since the CO objective involves minimization while the PG objective maximizes the reward by default.

\textbf{Training Policy Network}: To train the network, we set our objective function to maximize the terminal reward: 

\begin{equation}\label{eq:train_obj}
    J(\bt) = \mb{E}_{\tau \sim \pi_{\bt}}\left[R(\psi_{\tau}) \right] 
\end{equation}

We compute the gradient $\nabla J(\bt)$ using REINFORCE~\cite{williams1992simple} algorithm. However, since our reward structure is sparse and the action space is combinatorial we resort to ``greedy rollout'' baseline computation~\cite{kool2018attention} $b(s_t)$ to reduce the variance: 

\begin{equation}\label{eq:gradient_obj}
    \nabla_{\bt}J(\bt) = \mb{E}_{\tau \sim p_{\bt}} \left[\sum_{t=0}^{T} (R(\psi_\tau) - b(s_t)) \nabla_{\bt} \log \pi_{\bt}(a_t | s_t)\right]
\end{equation}

\subsection{Policy Network}
To compute the objective in Equation~\ref{eq:train_obj}, we implement a policy network with an encoder-decoder architecture. We describe the architecture of these components in detail below.

\subsubsection{Encoder}\label{sec:pencoder}
\begin{figure}[h]
    \centering
    \includegraphics[scale=0.8]{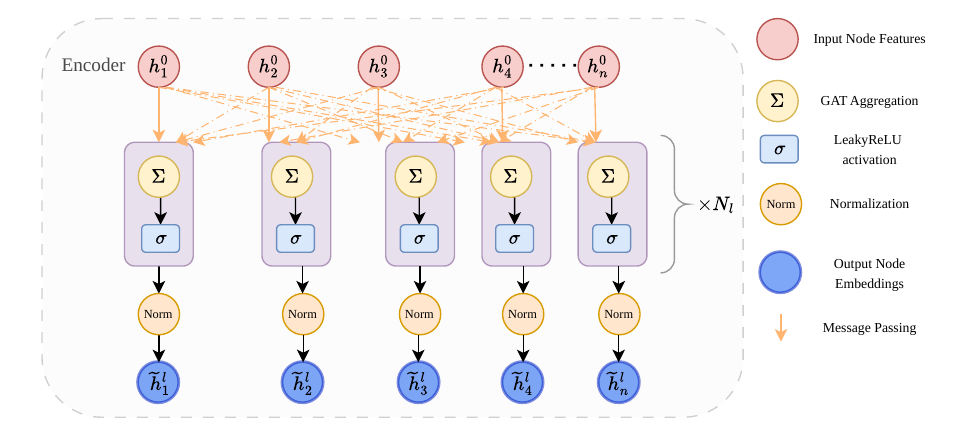}
    \caption{The policy network encoder accepts either a program or a coupling graph and produces node-level representations using a GNN encoder.}
    \label{fig:qmapping_encoder}
\end{figure}

The encoder showing in Figure~\ref{fig:qmapping_encoder} is responsible for computing the embeddings of the given $\Mc{P}_q$ and $G_Q$. Each $v_q \in \Mc{P}_q$ consists of node features $\bracks{\bm{x}_i}_{i=1}^{n}$. For some datasets, we preprocess quantum circuits to derive $\bm{x}_i \in \mb{R}^6$ features (Refer to Appendix~\ref{app:manual_feat_engg} for more details). For generic datasets, a one hot encoding $\bm{x}_i \in \mb{R}^n$ is a good initial choice. Nodes $V_Q \in G_Q$ also can have preprocessed features, although in this study we simply initialize $\bm{x}_Q^j \in \mb{R}^{N}$ with a one hot encoding. 

We denote $\tilde{\bh}_q$ as the node embeddings produced for $v_q \in \Mc{P}_q$ and $\hat{\bh}_Q$ for the node embeddings produced for $V_Q \in G_Q$. Both graphs are processed using an $N_{\ell}$ layer GAT~\cite{velivckovic2017graph} encoder: 
\begin{equation}\label{eq:gat_enc}
\tilde{\bh}_i^{(\ell)} = \text{Norm}\!\left(
  \overset{K}{\underset{k=1}{\Big\|}}\sigma\left(\alpha_{ij}^k\bm{W}^{k,(\ell)}h^{(\ell-1)}_j\right)\right) 
\end{equation}

Where, $\alpha_{ij}^k$ refers to the pairwise multi-head attention between two nodes $v_i, v_j$.  Additionally, $\text{Norm}$ is denotes a type of normalization. We performed experiments with LayerNorm~\cite{ba2016layer}, BatchNorm~\cite{ioffe2015batch} and GraphNorm~\cite{liu2019graph}. Out of these three, LayerNorm performs the worst both in terms of overall reward and generalization. We found little difference in performance with BatchNorm and GraphNorm. 
% TODO: recenter the diagrams
\subsubsection{Decoder}\label{sec:pdecoder}
\begin{figure}[t]
    \centering
    \includegraphics[scale=0.65]{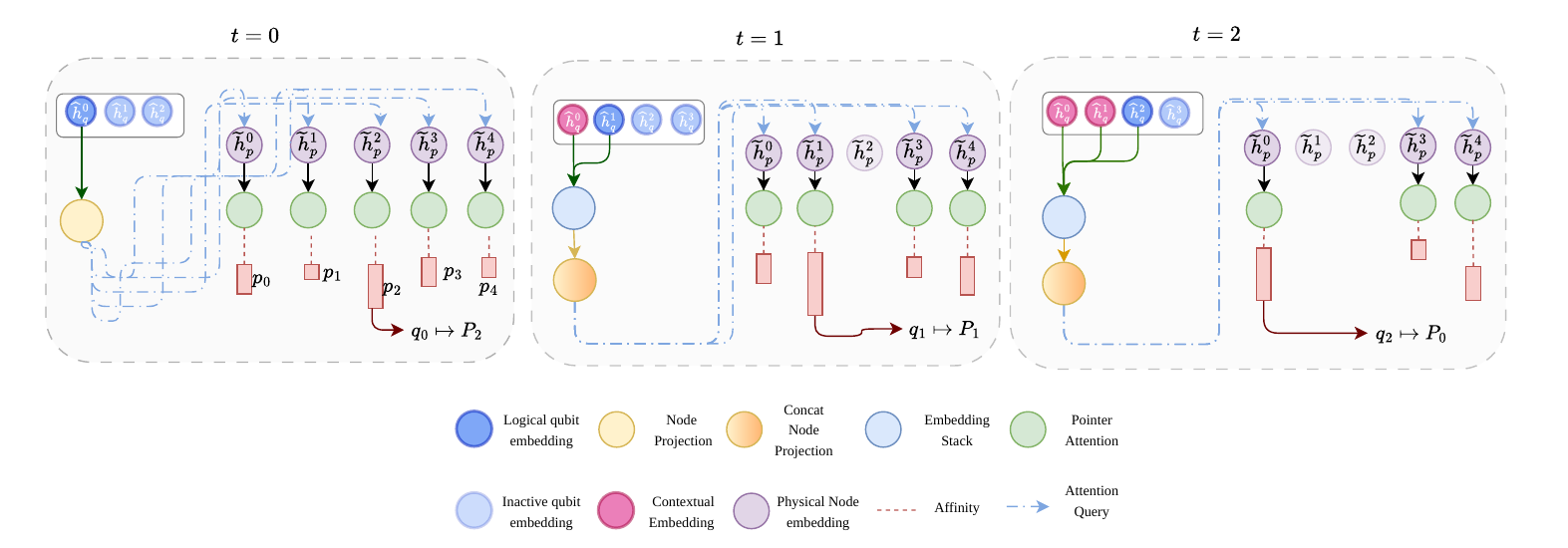}
    \caption{The decoder accepts program and coupling graph node embeddings from the encoder. At each decoding step, the selected nodes are added to the solution and those nodes are masked out (light shaded nodes) during the next step in decoding. At the termination of the process, the reward is calculated as the number of SWAPs introduced with the current solution.}
    \label{fig:qmapping_decoder}
\end{figure}

The embeddings $\bracks{\tilde{\bh}_i \in \mb{R}^{d_e}}_{i=1}^{n}$ and $\bracks{\hat{\bh}_j \in \mb{R}^{d_{e}}}_{j=1}^N$ where $d_e$ is the embedding dimension,  enable us to perform the iterative decoding step. An example of decoding for $t=0, 1, 2$ is shown in Figure~\ref{fig:qmapping_decoder}.

\textbf{Context Encoding}: Different from encoding $\Mc{P}_q, G_Q$ in a common embedding space, the context encoder is responsible for creating a fixed dimensional embedding $\bm{c} \in \mb{R}^{d_c}$ that serves as a \emph{query} for the decoder. The richer the context, the better the placement produced by the algorithm. We propose three different ways to compute the context in this work:

\begin{itemize}
    \item \textit{Project and Concat}: For the given current node $c_t$, we compute the previously placed logical qubit $p = c_{t-1}$. The projected embeddings are computed as $\tilde{\bh}_c' = \bW \tilde{\bh}_c$ and $\tilde{\bh}_p' = \bW \tilde{\bh}_p$ and $\bW \in \mb{R}^{d_e \times d_c/2}$. After projection, we derive $\bm{c} = [\bh_c';\bh_p']$. Here $[;]$ is a concatenation operation. 

    \item \textit{Concat and Project}: Instead of projecting twice, we compute $\tilde{\bh}' = [\tilde{\bh}_c;\tilde{\bh}_p]$ and then derive $\bm{c} = \bW \tilde{\bh'}$ and $\bW \in \mb{R}^{2d_e \times d_c}$. 

    \item \textit{Stack and Project}: In this case we perform the following context derivation:
    \begin{equation}\label{eq:stack_project_context}
        \bm{c} = \begin{cases}
            \bW \tilde{h}_c & t=0\\
            \bW \tilde{h}_p^{stack}  & t > 0
        \end{cases}
    \end{equation}
    Where $\tilde{h}_p^{stack} \in \mb{R}^{n_t \times d_e}$ is the stacked encoding. $n_t$ refers to the number of time steps up to (and including) the current timestep. Here $\bW \in \mb{R}^{d_e \times d_e}$
\end{itemize}

At each step during decoding we use the context encoding $\bm{c}$ as a query and compute $M$-head attention ($M > 1$) with $\hat{\bh}_Q$. We first project:
\begin{align}\label{eq:proj}
\bm{q}^{(c)}&=\bW_Q \bm{c};\quad\bk_i=\bW_K\hat{\bh_i};
\end{align}

Where $\bW_Q \in \mb{R}^{d_c \times d}$, $\bW_K \in \mb{R}^{d_e \times d}$. The attention score between the query and physical node embeddings in a single head (Pointer attention in Figure~\ref{fig:qmapping_decoder}) is calculated as: 
\begin{equation}\label{eq:attention}
   \mb{A}_{i}^{c} = C.\text{tanh}\left(\frac{\bm{q}^{(c)\top}\bk_i}{\sqrt{d}}\right)
\end{equation}
Following~\cite{bello2016neural}, we clamp the attention between $[-C, C]$ using $\text{tanh}$ nonlinearity after concatenating information from all $M$ heads. We set $C=10$ following existing work. This attention is then used to compute a masked probability distribution for selection the next action in Equation~\ref{eq:masked_qubit_policy}.

\subsection{Postprocessing Algorithm}
\begin{algorithm}
\caption{Post Processing Local Search}
\label{alg:rl_pp_alg}
\begin{algorithmic}[1]
\Require $\Mc{P}_q, G_Q$: Program and coupling graphs; $f_{S}: \mb{\Mc{A}} \mapsto \mb{R}$: SWAP computation function; $\Mc{N}_o$: Neighborhood op; $\pi_{\bt^*}$: The trained RL policy,  $N_{iters}$: Number of iterations;  $P$: Patience.

\Ensure $A_{best}$: Allocation s.t. $f_{S}(A_{best}) \leq f_{S}(A^{\pi^*})$ \Comment{Equality happens if no better assignments can be found.}

\State $C_{rl} \gets f_{S}(A^{\pi^*})$ \Comment{$\pi^* := \pi(\bt^*)$ and $A^{\pi^*}$ is the allocation produced by the current trained policy.}
\State $A_{best}, A_{curr} \gets A^{\pi^*}$
\State $C_{best}, C_{curr} \gets C_{rl}$
\State $p \gets 0$
\For{$i=1\dots N_{iters}$}
   \State $A_i \gets \Mc{N}_o(A_{curr})$ \Comment{Swaps after applying a local search op}
   \State $C_i \gets f_{S}(A_i)$
   \If{$C_i < C_{curr}$}
        \State $A_{curr} \gets A_i$
        \State $C_{curr} \gets C_i$
        \If{$C_{curr} < C_{best}$}
            \State $C_{best} \gets C_{curr}$
            \State $A_{best} \gets A_{curr}$
        \EndIf
    \Else
        \State $p \gets p + 1$       
   \EndIf
   \If{$p > P$}
    \State \textbf{break}; 
   \EndIf
\EndFor
\Return $A_{best}$
\end{algorithmic}
\end{algorithm}

Algorithm~\ref{alg:rl_pp_alg} outlines the algorithm based on local search~\cite{selman2006hill} to improve the quality of solutions proposed by the RL algorithm. It accepts $\Mc{P}_q, G_Q$ and the trained policy $\pi(\bt^*)$. To assist in the search we define two variants of the neighborhood operation function $\Mc{N}_o$: 

\begin{enumerate}
    \item \textit{Random Swap}: This operation selects an assignment $q_i \mapsto Q_k, q_j \mapsto Q_l \in A_{curr}$ and swaps them i.e. $q_j \mapsto Q_k ; q_i \mapsto Q_l$. 

    \item \textit{Random Assignment}: This operation first selects a random physical node $Q_k \in G_Q$. The node itself may already have an assignment or be unassigned. In the latter case, we select a random $q_i \in \Mc{P}_q$ and perform an assignment $q_i \mapsto Q_k$. In the former, case we swap the assignment as above. 
\end{enumerate}
 The key difference between the two methods is that we always find assigned physical qubit in the first method while the second can include unassigned (and potentially better) physical qubit assignments.

\section{Details of Training}
We train the policy network by generating random graphs with an edge probability $p=0.3$. The device graph topology is fixed for a particular policy network. All policy networks use an encoder with $4$ GATConv layers each with $8$ attention heads. The decoder uses $M=16$ multi-head attention heads. For all components (i.e. encoder, context encoder and decoder) we set embedding dimension $d=128$.
% To train the policy network described above, we generated random quantum programs. Since we do not care about single gate operations in this stage of our pipeline, it suffices to model quantum programs as DAGs where an edge $(u, v)$ denotes an interaction between qubits $u$ and $v$. The directionality stems from the fact that one of the qubits (called the control qubit) ``acts on'' the other qubit (called the target qubit). We generated these quantum programs with an edge likelihood $p=0.3$. Since quantum compilation is a device dependent task, we fix the topology of the device graph during training. 

The RL training is implemented with the \texttt{rl4co}~\cite{berto2025rl4co} library. We use a batch size of $512$ and a learning rate of $3e^{-4}$ with the Adam~\cite{kingma2014adam} optimizer.  During each episode, we consume the program qubits in a sequential manner for all program graphs in a batch. We only place the next qubit after the placement for the current qubit has been placed. However, we don't immediately compute the reward. The reward is computed at the end of an episode when all qubits in all program graphs have been placed according to Equation~\ref{eq:reward_func}. To train this model, we also use a ``rollout baseline''~\cite{kool2018attention} where during a given epoch $e$, we perform a rollout with policy $\pi_{\bt^e}$ on a randomly generated validation dataset.  The baseline is then the average reward produced over this dataset. We point out that even though in our training we sequentially consume qubits, there is no restriction on the order of consuming the qubits. It can be simple as ours to more complex like out-degree dependent scheduling. We leave such experiments for future work. We use a single NVIDIA H200 GPU to train the policy network. On average for most configurations, training lasts for $\sim 3$ hours.

We evaluate the trained policy network on data that is completely different than the data generated during training. More specifically, we intentionally evaluate the policy on data that is \emph{distributionally} different from the training set. This setup is essential to interpret results below because we cannot expect in-distribution data when the policy network is deployed in the compiler. The measure of performance is the number of SWAPs obtained after an allocation is obtained. Since one SWAP gate adds an overhead of 3 CNOT~\cite{siraichi2018qubit} gates, the more the number of SWAPs the greater the depth of the circuit and slower the performance. 

The test data in this works comes from three different datasets. The Munich Quantum Toolkit Benchmark (MQTBench)~\cite{mqt} is a dataset consisting of $\sim 70,000$ circuits with qubits ranging from $n=2$ to $n=72$.  The circuits themselves encompass various different quantum tasks like VQE~\cite{VQE}, QAOA~\cite{qaoa2014}, QPE~\cite{kitaev1995quantum, kitaev2002classical} etc. For circuits in this dataset, we perform manual feature engineering to extract initial features for the program graphs (see Appendix~\ref{app:manual_feat_engg}). Out of these we select circuits consisting of $15$ qubits as a good candidate for getting an estimate of the performance. We also consider the Queko benchmark~\cite{queko-tc20} for establishing the results on $n= \bracks{16, 20}$ qubits. The MQTBench test dataset consists of $166$ circuits, Queko-16 consists of $180$ and Queko-20 consists of $450$ circuits. To benchmark our method, we report the SWAPs obtained by Qiskit's SABRE~\cite{li2019tackling, zou2024lightsabre} compiler. To ensure a fair comparison, we isolate the \texttt{SabreLayout} and \texttt{SabreRouting} stages in the pipeline and report the results on the SWAPs produced by \emph{only} these stages.

For each dataset considered in the study, we perform decoding under four different decoding strategies~\cite{bello2016neural}. The decoding strategies dictate how the placement is selected from the likelihood distribution $p(a|s,c, \bt^*) = \pi_{\bt^*}(a_t|s_t, c_{<t})$. The greedy decoding strategy selects $a_t = \arg \max p(a|s,c, \bt^*)$. The sampling strategy selects an action $a_t \sim p(a|s,c, \bt^*)$. The other two strategies are multistart versions of greedy and sampling strategies. These strategies perform the same decoding but by generating $k > 1$ solutions and selecting the best performing one. We fix $k=10$ in the multistart decoding methods in our pipeline.

\begin{table}[t]
\centering
\caption{Average SWAPs introduced on a 64-qubit grid architecture with different settings of the policy network averaged over three independent runs with different seeds. The results are benchmarked against Qiskit's layout and routing stages. Red color indicates the best solution and blue indicates the second best solution obtained for the dataset.}
\label{tab:qubit_mapping_main_results}
\resizebox{\textwidth}{!}{%
\begin{tabular}{c|cccc|cccc|c}
\toprule
&  \multicolumn{4}{c|}{\textbf{RL}} & \multicolumn{4}{c|}{\textbf{RL + Post Processing}} & \textbf{Qiskit} \\
& Greedy & Sampling & \makecell{MultiStart\\Greedy} & \makecell{MultiStart\\Sampling} & Greedy & Sampling & \makecell{MultiStart\\Greedy} & \makecell{MultiStart\\Sampling} & \textbf{Mapping} \\
\midrule
% \rotatebox[origin=c]{90}{\textbf{Random}} &  & 29.44 & 28.73 & 28.31 & 29.50 & 1.29 & 1.20 & 1.21 & 1.29 & 174.11 \\[6pt]
\rotatebox[origin=c]{90}{\textbf{MQTBench}} & $88.82 \pm 2.0$ & $90.83\pm1.6$ & $96.24 \pm 1.4$ & $95.69 \pm 2.4$ & \color{red}$45.16 \pm 2.3$ & $51.19 \pm 5.2$ & $\color{blue}47.20 \pm 1.3$ & $45.81 \pm 2.1$ & $264.35$ \\[1.5pt]
\midrule
\rotatebox[origin=c]{90}{\textbf{Queko-16}} & $22.87 \pm 0.2$ & $24.42\pm 1.0$ & $24.82\pm 1.4$ & $23.94\pm 0.7$ & \color{blue} $0.15 \pm 0.02$ & $0.42 \pm 0.1$ & $0.25 \pm 0.04$ & $0.27\pm 0.03$ & \color{red}$0.116$ \\[1.5pt]
\midrule
\rotatebox[origin=c]{90}{\textbf{Queko-20}} & $52.74\pm 0.4$ & $53.09\pm 0.2$ & $53.11\pm 0.2$ & $52.71\pm 0.6$ & \color{blue}$6.14\pm 0.1$ & \color{red}$6.06\pm 0.3$ & $5.76\pm 0.1$ & $6.42\pm 0.2$ & $147.98$ \\[1.5pt]
% \multirow{2}{*}{\rotatebox[origin=c]{90}{\textbf{QUEKO}}} & $n=16$ & 22.87 & 24.42 & 24.82 & 23.94 & 0.15 & 0.42 & 0.25 & 0.27 & 0.116 \\[6pt]
% & $n=20$ & 52.74 & 53.09 & 53.11 & 52.71 & 6.14 & 6.06 & 5.76 & 6.42 & 147.98 \\[15pt]
\bottomrule
\end{tabular}%
}
\end{table}

% \begin{tabular}{c|cccc|cccc|c}
% \toprule
% & \multicolumn{4}{c|}{\textbf{RL}} & \multicolumn{4}{c|}{\textbf{RL + Post Processing}} & \textbf{Heuristic} \\
% & Greedy & Sampling & \makecell{MultiStart\\Greedy} & \makecell{MultiStart\\Sampling} & Greedy & Sampling & \makecell{MultiStart\\Greedy} & \makecell{MultiStart\\Sampling} & \textbf{Mapping} \\
% \midrule
% \rotatebox[origin=c]{90}{\textbf{MQTBench}} & -- & -- & -- & -- & -- & -- & -- & -- & -- \\[6pt]
% \midrule
% \rotatebox[origin=c]{90}{\textbf{Queko-16}} & -- & -- & -- & -- & -- & -- & -- & -- & -- \\[6pt]
% \midrule
% \rotatebox[origin=c]{90}{\textbf{Queko-20}} & -- & -- & -- & -- & -- & -- & -- & -- & -- \\[6pt]
% \bottomrule
% \end{tabular}
\section{Results}
Table~\ref{tab:qubit_mapping_main_results} summarizes the average SWAPs obtained by our algorithm on the three datasets for an $8\times8$ grid of $N=64$ physical qubits.  We report results obtained with and without post-processing (Algorithm~\ref{alg:rl_pp_alg}) on all decoding strategies.

On the MQTBench dataset with $n=15$, our method without any post-processing produces $\bm{63-66\%}$ less SWAPs than Qiskit's algorithm. With post-processing, the SWAPs are reduced by $\bm{80-82\%}$. On Queko-16, the RL algorithm performs worse than Qiskit but with post processing closely matches the number of SWAPs obtained by the algorithm. On Queko-20, the algorithm again outperforms Qiskit's algorithm by $\mathbf{64\%}$ without preprocessing and by $\mathbf{95\%}$ with post-processing.

\begin{figure}[h]
    \centering
    \includegraphics[width=\linewidth]{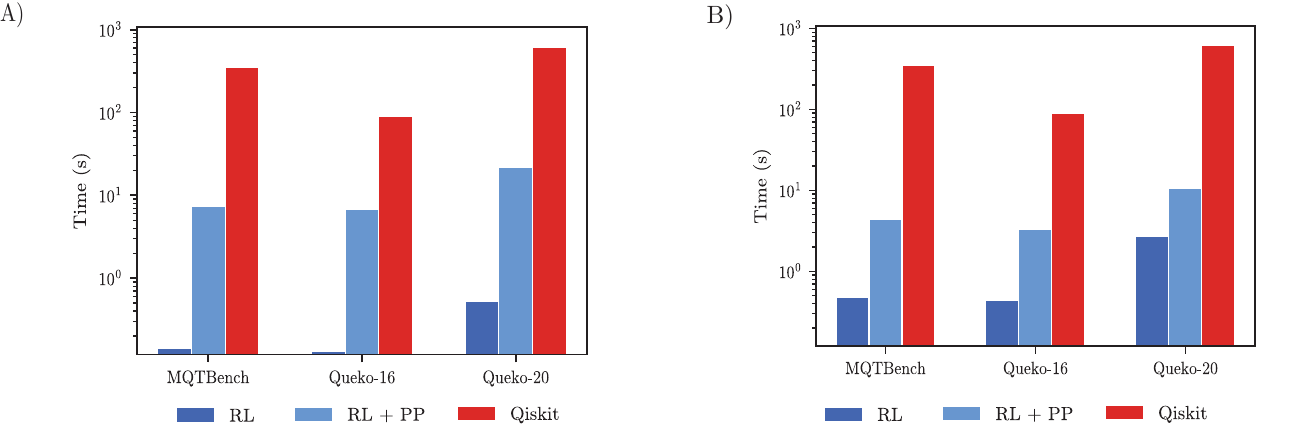}
    \caption{Time taken to produce mapping with our method vs Qiskit SABRE compiler. A) shows processing times when our pipeline uses a single GPU, B) shows processing times when only CPU is used.}
    \label{fig:inference_times}
\end{figure}

We further report the end-to-end wall clock time taken by our algorithm vs Qiskit's compiler in Figure~\ref{fig:inference_times}. Our results are obtained by running the policy network on the test set using a single NVIDIA H200 GPU and on a CPU, while Qiskit's results are reported on the CPU. This is because we do not possess a way to run the compilation pipeline of Qiskit on a GPU. The key takeaway from both the table and the plot is that our method is faster and produces considerably lower SWAPs in allocating qubits. We refer the reader to Appendix~\ref{app:ibm_hh_results} for a summary of results on the $65$ qubit IBM ``heavy hex''~\cite{paul2021ibm} architecture. For completeness, we also perform experiments with various optimization levels of the Qiskit compiler and benchmark the SWAPs against our pipeline in Appendix~\ref{app:qiskit_settings_ablation} and report the SWAPs produced for different types of quantum circuits in Appendix~\ref{sec:swaps_vs_circ_type}.
% and Appendix~\ref{app:qiskit_settings_ablation} for an ablation where we compare the SWAPs obtained with different optimization level settings of the SABRE compiler.

% We compare our results under four different decoding strategies~\cite{bello2016neural}. The greedy decoding strategy samples $\arg\max \pi_{\bt}(a|s)$. The sampling strategy samples a solution from the probability distribution $\pi_{\bt}(a|s)$. Multi-start methods also perform the same decoding but by generating $k > 0$ solutions and selecting the best performing one. We further benchmark the run time of our algorithm against Qiskit's layout and routing stages by reporting the end-to-end wall-clock time. Our method uses a single NVIDIA H200 GPU while Qiskit is CPU only. The wall-clock times are representative of algorithmic time under practical deployment. Our method (even factoring out the GPU) is considerably faster and generalizes to unseen data much better than Qiskit's algorithm. 

\subsection{Effect of Different Context Encodings}
\begin{table}[th]
\centering
\caption{Average SWAPs introduced by policy network trained with different context encoding styles on a $64$-qubit grid architecture with different context encodings on the MQTBench dataset with $15$ qubit circuits.}
\label{tab:ablation_ce}
\resizebox{\textwidth}{!}{
\begin{tabular}{c|c|c|c|c}
\toprule
   Encoding Strategy & Greedy & Sampling & Multistart Greedy & Multistart Sampling \\
\midrule
    Project and Concat & 97.41 & 93.68 & 85.86 & 85.35\\ 
    Concat and Project & 88.82 & 90.83 & 96.24 & 95.69\\
    Stack and Project & 133.60 & 126.79 & 135.97 & 146.04 \\ 
\bottomrule
\end{tabular}
}
\end{table}

One key component of our architecture is the way we generate context encodings during the decoding stage. Intuitively, the context provides a way for the policy network to ``see" the decisions made in previous time steps. Table~\ref{tab:ablation_ce} shows the average SWAPs obtained by context encoding method proposed in Section~\ref{sec:pdecoder}. From the table it becomes clear that Stack and Project context encoding is the worst performing of all schemes. For the other two, we see that ``project and concat'' performs better when the decoding methods are used with $k>1$ and ``concat and project'' works better when used with $k=1$ decoding.

\section{Conclusion}
NP Complete problems like qubit allocation typically are approached from either a heuristic perspective or application of RL to optimize a proxy metric without regard for the structure of the problem. In this paper, we show that an alternative approach exists  - if we can frame a given problem as a CO problem. Our work demonstrates that RL can outperform heuristic approaches when used in the right context. 

There are several avenues that open up from this work. For instance, the role of using other policy gradient algorithms like PPO~\cite{schulman2017proximal} has not been studied. We have also not used complex graph models like Graph PointerNets~\cite{yang2022graph} in this work and it will be interesting to see their use in this context. Finally,we assume that all device qubits are available with high uptime. In practice, this may strictly not be true. Training policy networks that can adapt to dynamic device conditions are left as avenues for future work. In addition, in our future work, we would also like to include the qubit routing stage into consideration, which would be beneficial to the whole compilation pipeline and make it more efficient.

\bibliographystyle{plain}
\bibliography{refs, ml_refs}

% space for appendices

\newpage
\appendix

\section{Pre-Processing Circuit Features for Qubit  Allocation}\label{app:manual_feat_engg}

Our algorithm does not depend on having access to input node features during placement. However, having access to good input program node features can definitely improve performance. In this section, we detail the feature engineering we perform on an input quantum circuit to generate node features for the program graph in the MQTBench dataset. These operations are however independent of the dataset and can be applied to any generic quantum circuit as well. 

We process the following features from a given quantum program:

\textbf{Single Operation density}: The single operation density $\mu_s^j$ is the ratio: 

\begin{equation*}
    \mu_s^j = \frac{\eta_s^j}{\eta}
\end{equation*}

Where $\eta_s^j$ is the number of single qubit gates on qubit $j$ and $\eta$ is the total number of gates.

\textbf{Entanglement Operation Density}: The entanglement operation density $\mu_m^j$ is the ratio: 
\begin{equation*}
    \mu_m^j = \frac{\eta_m^j}{\eta}
\end{equation*}

It is split into two components $\mu_c^j$ is the control qubit density and $\mu_t^j$ is the target qubit density. The former describes how many control operations are there on a particular qubit while the latter describes how many target operations exist on the qubit.

\textbf{Qubit Influence}: We represent a quantum circuit as a DAG. The influence score $\Mc{I}_j \in [0.0, 1.0]$ is a metric that quantifies how many qubits does the $j^{th}$ qubit ``influence'' at a $r$ distance random walk. 

\textbf{Pagerank Centrality Score}: We compute the \emph{pagerank}~\cite{page1999pagerank} centrality of a given qubit with a damping factor of $0.85$. The program graph is represented by $\tilde{\bm{A}} = \bm{D}^{-1}\bm{A}$ where $\bm{A}$ is the DAG adjacency matrix representation of the input quantum circuit and $\bm{D}$ is the node degree matrix.

\textbf{Quantum Causal Cone}: For each given qubit $q_i$ we compute it's quantum causal cone~\cite{leone2024practical} ($\Mc{C}_q^i)$. This is a conical region that contains all qubits that are affected by any change at $q_i$. Different from the influence score above, this metric returns the influence of a qubit in a fixed region. The score is computed as: 

\begin{equation*}
    S_{\Mc{C}} = \frac{\Mc{C}_q^i}{n}
\end{equation*}

\section{Benchmark Results on Other Topologies}\label{app:ibm_hh_results}
\begin{figure}[h]
    \centering
    \includegraphics[width=0.5\linewidth]{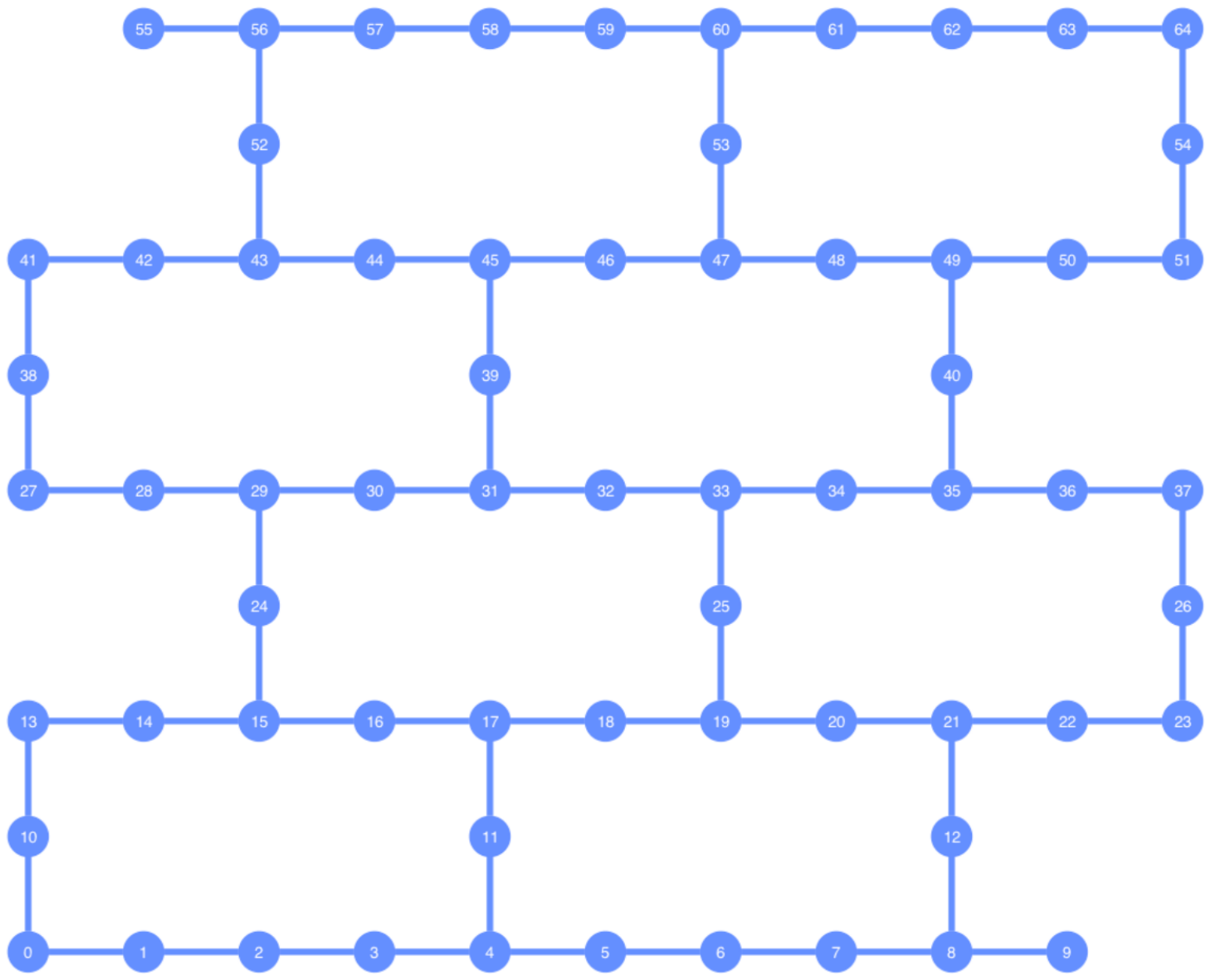}
    \caption{The 65-qubit IBM ``heavy hex'' lattice device graph topology.}
    \label{fig:ibm_hh_topology}
\end{figure}
Figure~\ref{fig:ibm_hh_topology} shows a IBM ``heavy hex'' topology consisting of 65 qubits. It has a sparse connectivity and thus it is much harder for algorithms to find a good layout. We train our policy network with ``concat and project'' contextual encoding while keeping the rest of the settings the same as above. Additionally, we do not employ any optimization in the SABRE layout and routing stages. 

\begin{table}[h]
    \centering
    \caption{Number of SWAPs obtained by our proposed algorithm on the IBM ``heavy hex'' devices. The baseline is the  default layout and routing methods provided by IBM's SABRE Compiler. The results are averaged over three seeds; standard deviations are similar to results presented above.}
    \label{tab:ibm_hh_results}
    \resizebox{\textwidth}{!}{%
    \begin{tabular}{c|cccc|cccc|c}
    \toprule
    & \multicolumn{4}{c|}{\textbf{RL}} & \multicolumn{4}{c|}{\textbf{RL + Post Processing}} & \textbf{Qiskit} \\
    & Greedy & Sampling & \makecell{MultiStart\\Greedy} & \makecell{MultiStart\\Sampling} & Greedy & Sampling & \makecell{MultiStart\\Greedy} & \makecell{MultiStart\\Sampling} & \textbf{Mapping} \\
    \midrule
    \rotatebox[origin=c]{90}{\textbf{MQTBench}} & 78.99 & 70.92 & 72.75 & 73.33 & 25.48 & 27.54 & 25.54 & 26.69 & 481.00 \\[6pt]
\midrule
\rotatebox[origin=c]{90}{\textbf{Queko-16}} & 13.64 & 14.51 & 14.38 & 13.87 & 0.02 & 0.00 & 0.03 & 0.03 & 273.66 \\[6pt]
\midrule
\rotatebox[origin=c]{90}{\textbf{Queko-20}} & 30.40 & 29.53 & 30.58 & 29.85 & 0.58 & 0.66 & 0.72 & 0.73 & 427.39 \\[6pt]
\bottomrule
\end{tabular}%
}
\end{table}

Table~\ref{tab:ibm_hh_results} summarizes the average SWAPs obtained on the heavy-hex topology. As before, our method with no post processing outperforms manual heuristic driven method by \textbf{85\%} on MQTBench dataset, \textbf{94\%} on Queko-16 and \textbf{92\%} on Queko-20 benchmark. With postprocessing the SWAPs are reduced by \textbf{94\%} on MQTBench, \textbf{99\%} on Queko-16, \textbf{99.8\%} on Queko-20. Remarkably, on this sparse architecture our policy outperforms Qiskit on Queko-16 without requiring post processing.
% Table~\ref{tab:ibm_hh_results} summarizes the average SWAPs obtained on a 65 qubit, hexagonal lattice (Figure~\ref{fig:ibm_hh_topology}). The topology has sparse connectivity and is thus a harder instance of the topology we consider in the main result. The policy networks are trained with a ``concat and project'' contextual encoding. All other settings are unchanged from the main results. 

% \section{Benchmark Results against Greedy Mapper}

\section{Ablation with Qiskit SABRE Optimization Levels}\label{app:qiskit_settings_ablation}

\begin{figure}[h]
    \centering
    \includegraphics[width=\linewidth]{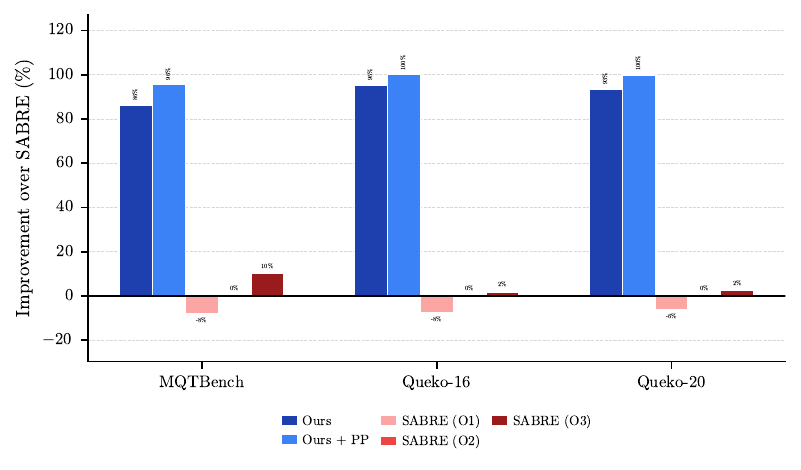}
    \caption{Improvements in SWAPs obtained with our solver and various optimization level settings for the Qiskit SABRE Compiler on the 65 qubit IBM Heavy-Hex Architecture.}
    \label{fig:ibm_hh_sabre_ablation}
\end{figure}

\begin{table}[h]
    \centering
    \begin{tabular}{c|c|c|c}
    \toprule
     Optimization Level    &  $L_{tr}$ & $S_{tr}$ & $N_{max}$ \\
    \midrule
     \textbf{O1} & 5 & 5 & 2 \\
     \textbf{O2} & 20 & 20 & 4 \\ 
     \textbf{O3} & 50 & 50 & 6\\ 
     \bottomrule
    \end{tabular}
    \caption{Optimization Level Settings for the SABRE Compiler}
    \label{tab:opt_levels}
\end{table}

To demonstrate that we do not cherry pick our results, we perform an ablation in which we vary the settings available to us in the SABRE compiler. Specifically, we vary the the number of \texttt{layout trials} ($L_{tr})$, \texttt{swap trials} ($S_{tr})$ and \texttt{max iteration}$(N_{max})$ settings in \texttt{SABRELayout} call. We consider three different optimization levels as shown in Table~\ref{tab:opt_levels}. The results on the IBM Heavy-Hex Topology~\ref{fig:ibm_hh_topology} are shown in Figure~\ref{fig:ibm_hh_sabre_ablation}. 

The results report the improvement in SWAPs over baseline. Our baseline here is the \textbf{O2} optimization setting in SABRE Compiler. We can see that a higher optimzation setting leads to a $2 - 10\%$ improvement over the baseline while \textbf{O1} setting is $6-8\%$ worse than baseline. The RL method (``Ours'') obtains a $86 - 96\%$ improvement over the baseline and with post processing (``Ours + PP'') leads to a $100\%$ improvement over the baseline.

\section{SWAPs Produced for Different Types of Circuits}\label{sec:swaps_vs_circ_type}

\begin{figure}[t]
    \centering
    \includegraphics[width=\linewidth]{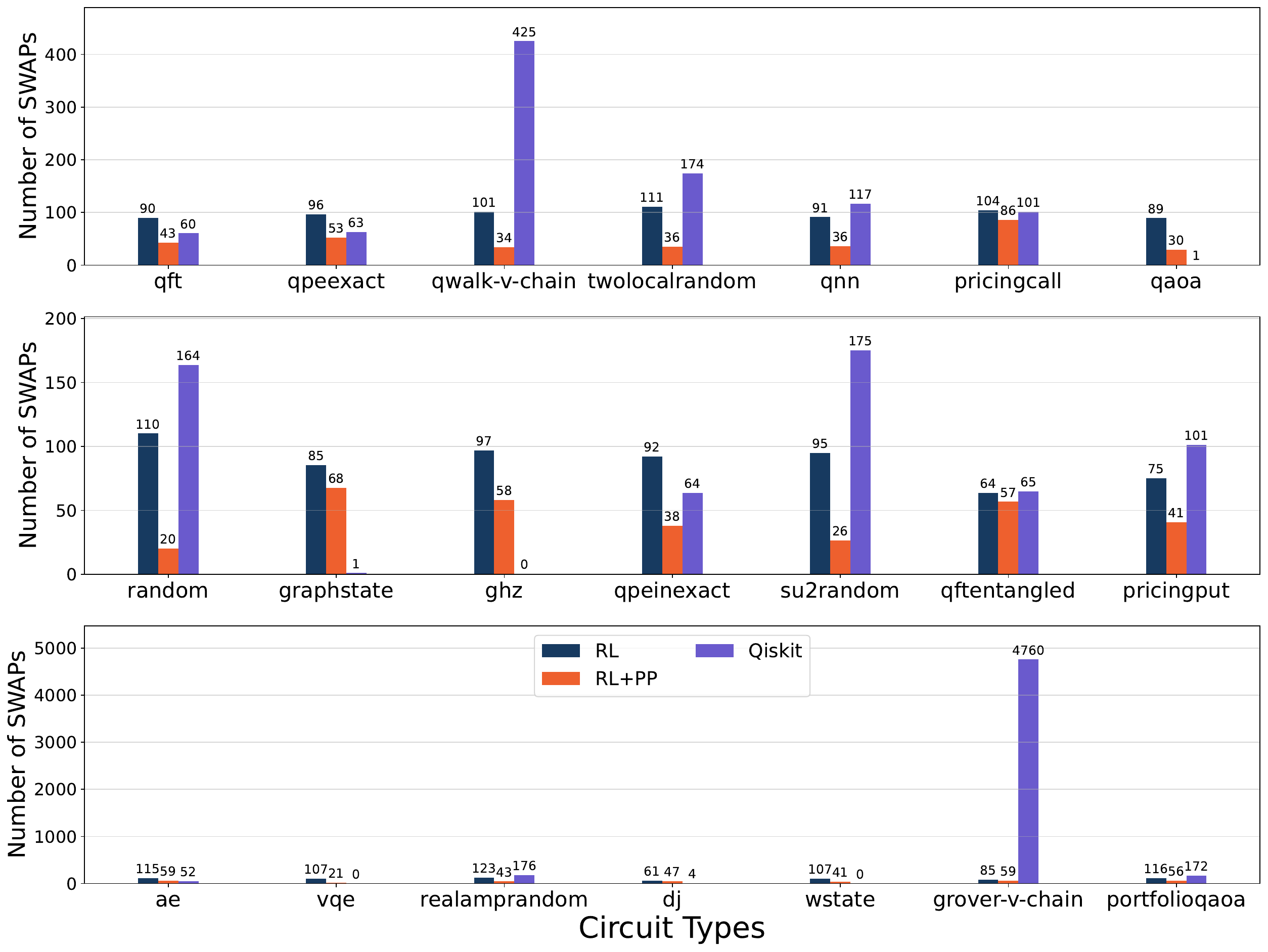}
    \caption{Number of SWAPs produced by our proposed method vs Qiskit SABRE compiler baseline for different types of quantum circuits. Our methods (RL, RL+PP) significantly outperform the SABRE compiler on most circuit types.}
    \label{fig:swaps_circ_types}
\end{figure}

\begin{figure}[h]
    \centering
    \includegraphics[width=\linewidth]{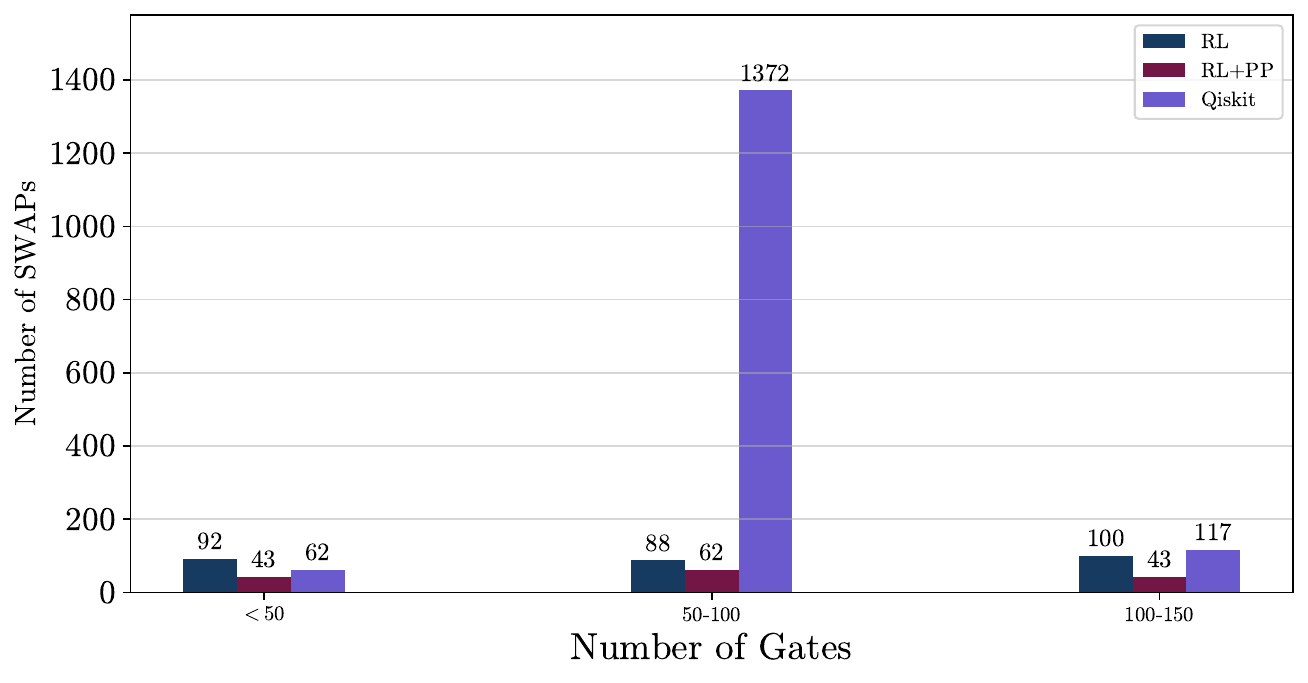}
    \caption{Number of SWAPs produced for different number of CNOT gates in program qubits.}
    \label{fig:swaps_vs_gates}
\end{figure}
In order to benchmark how our method performs on different types of quantum circuits, we gathered results on the MQTBench dataset for $n=15$ circuits and a $8\times8 (N=64)$ grid topology. The circuits are divided into $21$ different categories corresponding to a particular quantum computing task. We benchmark the SWAPs produced by the policy network (RL) and the SWAPs obtained after post-processing (RL+PP) against the Qiskit SABRE compiler with the default settings. Figure~\ref{fig:swaps_circ_types} shows the results of our benchmark. 

We can see that our methods (RL, RL+PP) outperform the SABRE Compiler in 18 out of 21 categories. One instance that we highlight is the significant ($97-98\%$) reduction obtained over the Qiskit SABRE method in the \texttt{grover-v-chain} type circuit. The results are further evidence that our method performs well across a variety of circuits as opposed to performing exceptionally well on one type of circuit (e.g. QAOA). 

Figure~\ref{fig:swaps_vs_gates} shows the number of SWAPs obtained by our algorithm on the MQTBench datasets compared to the number of input CNOT gates in the program graph. While Qiskit outperforms the RL method on circuits with less than $50$ gates, their scalability drastically worsens as the number of gates in the program graph increases. This implies that heuristic based methods may not be a good fit when scalability is considered. On the other hand, our trained RL policy introduces less than 100 SWAPs for circuits upto 100 CNOT gates. Even for larger circuits, with 100-150 gates, the trained policy produces 100 SWAPs on average. The best performance is achieved when post-processing is applied to the output of the RL policy. In this case, we achieve less than 100 SWAPs any quantum circuits with gates ranging from 0-150.

% It becomes clear from the figures that for of circuits our methods (i.e. RL, RL+PP) significantly outperform the SABRE compiler. Specifically, for \texttt{grover-v-chain} type circuits, out method obtains a stunning $97-98\%$ reduction in the number of SWAPs compared to 

% \input{checklist}

\end{document}